\documentstyle[preprint,eqsecnum,aps]{revtex}
\begin{document}
%
%
  \input epsf
%
\begin{titlepage}
\title{Time and Time Functions\\
 in Parametrized Non-Relativistic Quantum Mechanics}

\author{James B.~Hartle\thanks {e-mail: hartle@cosmic.physics.ucsb.edu}}
\vskip .13 in
\address{Department of Physics, University of California,\\
Santa Barbara, CA 93106-9530}

\draft
\preprint{UCSBTH-95-2}

\maketitle

\begin{abstract}
\tighten

The ``evolving constants'' method of defining the quantum dynamics of
 time-reparametrization-invariant theories is investigated for
a particular implementation of
parametrized non-relativistic quantum mechanics (PNRQM).  The wide
range of
time functions that are available to define evolving
constants raises issues of interpretation, consistency, and the degree to
which the resulting quantum theory coincides with, or generalizes, the usual
non-relativistic theory.  The allowed time functions must be
restricted for the predictions of PNRQM to coincide with those of usual
quantum theory.  They must be restricted to have a notion of quantum
evolution in a time-parameter connected to spacetime geometry.  They
must be restricted to prevent the theory from making inconsistent
predictions for the probabilities of histories.  Suitable restrictions
can be introduced in PNRQM but these seem unlikely to apply to a
reparametrization invariant theory like general relativity.

\end{abstract}
\end{titlepage}

\tighten

\narrowtext
\setcounter{footnote}{0}
\section{Introduction}
\label{sec:I}

Relativistic theories of gravity such as general relativity or string
theory are invariant under reparametrizations of time. The quantization
of such theories presents a number of problems of principle loosely
known as ``the problem of time''\footnote{For recent and, by now,
classic reviews of the problem of time see Refs
\cite{Ish93,Kuc92,Unr91}}. Beyond any issues of finiteness or
consistency such theories may exhibit, the usual quantum mechanical
framework for prediction must be formulated anew.
That is because a fixed notion of time is central to these usual frameworks ---
but no such fixed notion is supplied by a theory which does not
distinguish between two time variables differing by a reparametrization.

The method of ``evolving constants'' is a concrete proposal for defining
quantum dynamical predictions in the context of Dirac quantization of
theories with a single time reparametrization invariance. Features of
this idea may be found in the work of DeWitt \cite{DeW62}, Page and
Wootters \cite{PW83,Woo86}, and Carlip \cite{Car90a,Car90b}.
However, the most complete development is to be found in the work of Rovelli
\cite{Rov90a,Rov91} which we shall rely on.
We shall review the method of ``evolving constants'' in some
detail in Section III, but the essential features are as follows:

Classically a reparametrization invariance of the action implies a
constraint on the canonical co\"ordinates $q^i$ and their conjugate
momenta $p_i$ which may be written
\begin{equation}
H(p_i,q^i)=0\ .
\label{oneone}
\end{equation}
In a time-reparametrization theory $H$ is the superhamiltonian and may
be
chosen to generate reparametrizations.

In the Dirac approach to the quantization of constrained systems, the
$p$'s, $q$'s, and $H$ become operators acting on a linear space. The
operator $\widehat H$ annihilates physical states and commutes with
observables
\begin{equation}
[\widehat F,\widehat H]=0\ .
\label{onethree}
\end{equation}
(We use a caret to denote operators while quantities without carets are
classcal functions.) Observables are thus ``constants'' in the sense of
commuting with the superhamiltonian.

Eq.~(\ref{onethree}) would seem to
prohibit any reparametrization-invariant quantum evolution.  However,
the proposal of \cite{Rov90a,Rov91} is to define evolution using families of
operators labeled by a time parameter $\tau$.  Each operator in a
family satisfies (\ref{onethree})
and represents a physical
quantity at a value of $\tau$.
Each one-parameter family is generated from a time function ${\cal
T}(p_i,q^i)$.
Given a classical function on phase space $F(p_i,q^i)$, its value may be
found at an intersection of a classical phase space trajectory and the
surface ${\cal T}(p_i,q^i)=\tau$. There are different trajectories for
different initial conditions $(p_i,q^i)_0$,  at say $\tau=0$, and thus
the values $F$ at
$\tau$ become functions of $(p_i,q^i)_0$. When, with suitable
ordering, these functions of $p$'s and $q$'s are turned into operators,
there results a one parameter family of operators $\widehat F(\tau)$,
each satisfying
(\ref{onethree}), in which $\widehat F(\tau)$ may be said to represent the
classical quantity $F(p_i,q^i)$ at the time parameter $\tau$. This family
of ``evolving constants'' $\widehat F(\tau)$ may then be employed,
as in the usual
Heisenberg picture, to define evolution as a function of a time
parameter $\tau$. For example, one can discuss the variation in an
expected value $\langle \widehat F(\tau)\rangle$ as a function of a time
parameter $\tau$. Different time functions may be used to construct
different families of ``evolving constants''. Any time function in a
large set yields a family of evolving constants for each classical
quantity $F(p_i, q^i)$. The quantum mechanical probabilities that  result from
all families are the predictions for quantum evolution.

The severe difficulties that would arise in actually implementing such a
program for realistic systems have been discussed by H\' aj\'\i\v cek
\cite{Haj91}, Unruh \cite{Unr91}, and Kucha\v r
\cite{Kuc92}.
H\' aj\'\i\v cek and Kucha\v r stressed the
ambiguities that arise in finding a consistent operator ordering when
deriving quantum evolution from classical evolution.  Unruh emphasized
the serious problem with even exhibiting the classical evolution of a
chaotic system.  However, it is not the purpose of this paper to
reexamine these problems.\footnote{Recently A.~Anderson \cite{And95}
 has argued
that they may not be as severe as they seem.}
 Rather we shall assume that the difficulties
raised by H\'aj\'\i\v cek, Kucha\v r,
and Unruh can be solved and investigate a series of
complementary questions bearing on the consistency of
the ``evolving constant'' proposal and the extent to which it
reproduces the predictions of
familiar quantum mechanics in a fixed flat background spacetime.

Time reparametrization invariance is not just a feature of geometric
theories of gravity. As stressed by Dirac \cite{Dir64}
 and Kucha\v r \cite{Kuc74}, {\it any}
classical theory may be made time reparametrization invariant by
parametrizing the time and elevating it to the status of a canonical
variable. For example, suppose the dynamics of a single
non-relativistic particle moving in one dimension is summarized by an
action
\begin{equation}
S[x(t)] = \int\nolimits^{t^{\prime\prime}}_{t^\prime} dt\,\ell
\bigl(dx/dt, x\bigr)\ ,
\label{onefour}
\end{equation}
for some Lagrangian $\ell$. Simply by writing $t=t(\lambda)$ and
$x=x(\lambda)$ (``parametrizing'' them) this action may be rewritten as
\begin{equation}
S[x(\lambda),t(\lambda)] = \int d\lambda\, \dot t\ell \bigl(\dot x/\dot
t, x\bigr)\ ,
\label{onefive}
\end{equation}
where a dot denotes a derivative with respect to $\lambda$. This is an
equally good summary of classical dynamics, for the equations of motion of
(\ref{onefive}) imply those of (\ref{onefour}). The action
(\ref{onefive}) is evidently reparametrization invariant because the
way in which $x$ and $t$ were parametrized was not specified in its
construction.  This parametrized mechanics is
fully equivalent to usual mechanics.

A quantum theory of parametrized non-relativistic mechanics (PNRM) may be
constructed by restricting $\ell$ to be of usual non-relativistic form,
employing the principles of Dirac quantization, and calculating quantum
evolution by the method of ``evolving constants''. To what
extent does the resulting parametrized non-relativistic quantum
mechanics (PNRQM) coincide with familiar non-relativistic quantum
mechanics or differ from it? As shown in Refs \cite{Rov90a,Rov91},
with the choice of
time function
\begin{equation}
{\cal T} \bigl(p_x,p_t,x,t\bigr)=t
\label{onesix}
\end{equation}
the operators $\widehat F(\tau)$ coincide with the usual Heisenberg picture
operators.  The
predictions of PNRQM thus agree with usual non-relativistic quantum
mechanics for this choice of time function.

However, the method of evolving constants permits a much wider variety
of time functions than (\ref{onesix}), and the predictions
of all of them must be considered.  In this paper we discuss the character of
the predictions of PNRQM that arise from time functions other
than (\ref{onesix}). First we show  that the general time functions
yield predictions that are nothing like those of
non-relativistic quantum mechanics.  Next we point out that measurements
of  the time variables
corresponding to certain time functions are very different in character
from measurements of
the Newtonian time of non-relativistic spacetime. Finally, we
examine the predictions of the ``evolving constants'' method for time
histories.

Usual quantum mechanics predicts probabilities for sequences of
alternatives that constitute time histories.  Such probabilities are
necessary if the theory is to make predictions about such everyday
phenomena as the orbit of the moon or the evolution of the universe. In
predicting such probabilities, usual quantum mechanics employs, in
addition to the usual law of unitary evolution, some variant of a
``second law of evolution'' sometimes called ``the reduction of the
state vector''.

The original work on ``evolving constants'' did not explicitly address
the prediction of the probabilities for histories, but the usual quantum
mechanical rules for these predictions can be straightforwardly
included. However, we show in a parametrized relativistic field theory
that allowing arbitrary time functions in this straightforward extension
can lead to inconsistent predictions for the probabilities of field
histories.

In Section II we review what we mean by usual non-relativistic quantum
mechanics to which PNRQM is to be compared.  Section III reviews the
details of the ``evolving constants'' method for defining quantum
evolution.  In Section IV we compare the predictions of ``good'' time
functions which define surfaces in phase space that a classical trajectory
intersects at most once with ``bad'' time functions which do not have
this property.  Section V discusses the connection between time
functions and non-relativistic spacetime. In Section VI we display the
inconsistencies that may arise if arbitrary time functions are permitted
in theories with ``many-fingered time''. Section VII offers some brief
conclusions and opinions.

\section{Usual Non-Relativistic Quantum Mechanics}
\label{sec:II}

In this section we describe what we mean by usual non-relativistic
quantum mechanics.  We shall define this as narrowly and as
specifically as possible, not to be contentious concerning the meaning
of the term,
but to highlight any contrast with PNRQM.

We work in the Heisenberg picture where states are fixed
 vectors in a Hilbert space
${\cal H}$. An exhaustive set of ``yes-no'' alternatives at time $t$
is represented by a set of projection operators
$\{\widehat P_\alpha(t)\}$ where
the discrete index $\alpha$ labels the particular alternative.  These
operators evolve according to the Heisenberg equations of motion
\begin{equation}
\widehat P_\alpha(t) = e^{i\hat ht} \widehat P_\alpha(0) e^{-i\hat ht}\ ,
\label{twoone}
\end{equation}
where $\hat h$ is the Hamiltonian. (Here, as throughout, we employ units where
$\hbar=1$.)

A set of alternative time histories for the system is defined by a
succession of alternatives $\left\{\widehat P^1_{\alpha_1}(t_1)\right\},
\left\{\widehat P^2_{\alpha_2}(t_2)\right\}, \cdots$ at a sequence of times
$t_1
< \cdots < t_n$. An individual history corresponds to a
particular sequence of alternatives $\alpha_1, \cdots, \alpha_n$ whose
probability is
\begin{equation}
p\left(\alpha_n,\cdots,\alpha_1\right) = \left\Vert \widehat
P^n_{\alpha_n}(t_n)
\cdots \widehat P^1_{\alpha_1} (t_1) | \psi \rangle\right\Vert^2\ ,
\label{twotwo}
\end{equation}
where $|\psi\rangle$ is the Heisenberg state of the system. The
operators in (\ref{twotwo}) are time ordered. The
relation (\ref{twotwo}) exhibits compactly the two laws of evolution in
quantum mechanics --- unitary evolution between alternatives
[eq.\ (\ref{twoone})] and reduction of the state vector (by the action of
the projections) at them.

For the present discussion, (\ref{twotwo}) may be regarded in either of
two ways.  It can be thought of as the probability of the outcomes of a
sequence of ideal measurements on a subsystem in the approximate quantum
mechanics of measured subsystems (aka the ``Copenhagen'' formulation). In
that case ${\cal H}$ is the Hilbert space of the measured subsystem.
More fundamentally (\ref{twotwo}) may be thought of as the probability
of a history in a decohering set of alternative histories of a closed
system, most generally the universe. The reason (\ref{twotwo}) may be
viewed either way is that we shall only be concerned with whether PNRQM
reproduces the form of (\ref{twotwo}), not issues of decoherence.

While not always stated explicitly, usual non-relativistic quantum
mechanics assumes fixed Newtonian spacetime geometry.  In particular the
$t$ in (\ref{twotwo}) is the familiar Newtonian time.  That assumption
means that the time $t$ may be measured by a clock that is entirely
separate from any subsystem under study.  Indeed, that is how time is
measured in typical experiments.  It is because $t$ is a property of an
assumed fixed spacetime geometry, and not a property of the quantum
system itself, that it is represented in the theory as a parameter
describing evolution and not as an operator in the Hilbert space of the
quantum system.

\section{Parametrized Non-Relativistic Quantum Mechanics}
\label{sec:III}

In this section we lay out in more detail than was possible in the
Introduction the essential features of PNRQM \cite{Dir64,Kuc74}.
We shall be brief because we are merely reviewing the development of
\cite{Rov90a,Rov91}. In this and the succeeding two sections we restrict
attention to a non-relativistic particle moving in one dimension with a
Hamiltonian of the form
\begin{equation}
h(p_x,x) = \frac{p^2_x}{2m} + V(x)\ .
\label{threeone}
\end{equation}
We shall consider aspects of field theory in Section VI.

The extended phase space for the parametrized non-relativistic mechanics
summarized by the action (\ref{onefive}) is spanned by co\"ordinates
$(t, x)$ and momenta $(p_t,p_x)$. We write these $x^\mu$ and $p_\mu$
respectively where $\mu = 1,2$, and as $z^A$ when we wish to speak of all
four together. The $z^A$ are the co\"ordinates of a point in phase
space, $A=1,2,3,4$. The action of parametrized non-relativistic mechanics
(\ref{onefive}) implies
a constraint of the form
\begin{equation}
H\left(p_\mu, x^\mu\right) = p_t + h\left(p_x, x\right) =0\ ,
\label{threetwo}
\end{equation}
which defines the ``constraint surface'' in the extended phase space.
The constraint generates classical trajectories $\gamma$ in the
constraint surface according to
\begin{equation}
\dot z^A(\lambda) = \left\{z^A(\lambda), N(\lambda)  H\right\}\ ,
\label{threethree}
\end{equation}
where $\{~,~\}$ is the Poisson bracket and
$N(\lambda)$ is a multiplier (the ``lapse'') defining the
particular parametrization of these trajectories.
The set of classical trajectories $\gamma$ constitute the reduced phase
space. The functions $F(z^A)$ that are constant on these curves are the
classical observables and satisfy
\begin{equation}
\{F,H\}=0\ .
\label{threefour}
\end{equation}

To pass to quantum mechanics, we consider the linear space of wave
functions $\psi(x^\mu)$ and represent $x^\mu$ and $p_\mu$ by operators
in the usual way, {\it e.g.}, $p_\mu = -i\partial/\partial x^\mu$. Physical
states are annihilated by the operator form of the constraint
\begin{equation}
\widehat H\psi(x^\mu) = \left[-i\frac{\partial}{\partial t} +
\hat h\Bigl(-i\frac{\partial}{\partial x}, x\Bigr)\right] \psi(t,x) =0\ ,
\label{threefive}
\end{equation}
which will be recognized as the Schr\"odinger equation.  The inner
product between physical states is
\begin{equation}
(\psi, \phi) = \int\nolimits^{+\infty}_{-\infty} dx\, \psi^* (t, x)\phi
(t,x)
\label{threesix}
\end{equation}
and is independent of $t$ as a consequence of (\ref{threefive}).

To define a family of ``evolving constants'' we choose a time function
${\cal T}(z^A)$ such that every classical trajectory intersects each
surface of constant ${\cal T}$ at least once. (We shall return below to
the question of whether classical trajectories may intersect such
surfaces more than once.) Evidently ${\cal T}$ cannot be an observable
because then it would be constant along classical trajectories.

Consider any function $F(z^A)$ on the extended phase space. For each
value of a parameter $\tau$, we may define the classical observable
$F(\tau, \gamma)$ as having the value of $F(z^A)$ at the point $z^A$ where
$\gamma$ intersects the surface ${\cal T}(z^A) = \tau$. If more than one
intersection is possible then more than one family of observables
$F(\tau,\gamma)$ may be defined.

Classical trajectories may be labeled by their location $z^A_0$ in phase space
at $\lambda=0$. Thus $F(\tau,\gamma)$ becomes a function of
$z^A_0$ The idea now is to replace $x^\mu_0$ and $p_{\mu 0}$ by their
corresponding operators and carry out a suitable operator ordering to
yield a one parameter family of quantum observables $\widehat F(\tau)$
satisfying
\begin{equation}
\bigl[\widehat F(\tau), \widehat H\bigr]=0\ .
\label{threeseven}
\end{equation}
The operator $\widehat F(\tau)$ represents the quantity $F$
at the time parameter $\tau$.

Probabilities may be calculated as in ordinary quantum mechanics treating
the parameter $\tau$ as time. Suppose, for example, a sequence of ideal
measurements is made on a subsystem of ranges of values of quantities
$F^1,\cdots, F^n$ at a sequence of time parameters
$\tau_1 <\cdots <\tau_n$. Let $\hat P^k_{\alpha_k}(\tau_k)\, ,\, \alpha_k =
1,2,\cdots$ be the projection operators on the ranges of the spectrum of
the operator $\hat F^k(\tau_k)$ that define the possible outcomes of the
measurement at time parameter $\tau_k$.
Then the probability of a particular sequence of outcomes
$(\alpha_1,\cdots,\alpha_n)$ is
\begin{equation}
p\left(\alpha_n,\cdots,\alpha_1\right) = \left\Vert \widehat P^n_{\alpha_n}
(\tau_n)\cdots \widehat P^1_{\alpha_1} (\tau_1) |\psi\rangle\right\Vert^2\ ,
\label{threeeight}
\end{equation}
where the operators are ordered by the value of $\tau$.
A notationally indistinguishable formula holds for the probability of a
decoherent set of histories of a closed system. The probabilities of
a history of alternatives at a sequence of times were not explicitly
discussed in \cite{Rov90a,Rov91}.  In employing formulae such as
(\ref{threeeight}) we are assuming that histories with general time functions
would be treated just as they are in the Newtonian time.

With perhaps some suitable general restrictions, {\it any} choice of time
function yields a family of evolving constraints for each classical
quantity $F(p_i, q^i)$. For each such choice of time parameter, PNRQM
predicts probabilities according to (\ref{threeeight}). The totality of
all these probabilities for all allowed time functions are the
predictions of the quantum theory.

As discussed in Ref.~\cite{Rov90a,Rov91}, one time function
for which this procedure may be carried out explicitly is
\begin{equation}
{\cal T} (z^A) = t\ .
\label{threenine}
\end{equation}
For the one parameter of observables representing position and momentum
when ${\cal T}= t = \tau$ we  may take
\begin{mathletters}
\label{threeten}
\begin{eqnarray}
\hat x(\tau)& = &e^{i\hat h(\tau-\hat t)} \hat x\, e^{-i\hat h(\tau-\hat
t)}\ ,
\label{threetena} \\
\hat p_x(\tau)& = &e^{i\hat h(\tau-
\hat t)}\hat  p\, e^{-i\hat h(\tau-\hat t)}\ .
\label{threetenb}
\end{eqnarray}
\end{mathletters}
where the operators $\hat x$ and $\hat t$ act on wave functions $\psi(x^\mu)$
by
multiplication, and $\hat p$ is $-i\partial/\partial x$. The operators
(\ref{threeten}) commute with the constraint (\ref{threetwo}) and
classically correspond to the values of $x$ and $p_x$ when ${\cal T} =
t= \tau$. The value of any function $F(z^A)$ when ${\cal T}=\tau$ may be
similarly represented.

The quantities $\hat x(\tau)$ and $\hat p_x(\tau)$ are the usual
representations of
position and momentum in the Heisenberg picture.  Thus, for the choice
of time function ${\cal T}=t$, the predictions of PNRQM coincide exactly
through (\ref{threeeight}) with those of usual non-relativistic quantum
mechanics (\ref{twotwo}). However, PNRQM permits many more choices of
time functions, and it is to the predictions of these that we now turn.

\section{Good and Bad Time Functions}
\label{sec:IV}

In this section we consider some simple examples of time functions of
the form ${\cal T}={\cal T} (t,x)$ and show that they lead to
predictions which are not contained in usual non-relativistic quantum
mechanics.

We begin with the elementary example of a free particle, $h=p^2/2m$, and
investigate the time function ${\cal T} = x$. Almost all classical trajectories
intersect a surface $x=\tau$ once and only once. (The only exceptions are
those with zero momentum that may remain at $x=\tau$.)  The operators
corresponding to the values of the phase space co\"ordinates when
$x=\tau$ are
\begin{mathletters}
\label{fourone}
\begin{eqnarray}
\hat x(\tau) &=& \tau\ , \label{fouronea}\\
\hat p_x(\tau)& = & \hat p\ , \label{fouroneb}\\
\hat t(\tau) &=& \hat t + \frac{1}{2}[\frac{m}{\hat p},
(\tau-\hat x)]_+
\ , \label{fouronec} \\
- \hat p_t(\tau)&=& \hat p^2/2m\ . \label{fouroned}
\end{eqnarray}
\end{mathletters}
where $[~,~]_+$ is the anti-commutator making the operator symmetric.

Thus, we could calculate a probability for a measurement of $t$ at a
given value of $x$. For typical wave functions $\psi(x^\mu)$, the
expected value $\int dx \psi^* \hat t(\tau)\psi$ at a given value of
$x=\tau$ would diverge because of the
$1/\hat p$ factor in (\ref{fouronec}) but for wave functions with zero
amplitude for $p=0$ it might be finite.  Using
(\ref{threeeight}) one could calculate the probabilities for values of
$\hat t$ at a sequence of $x$'s. The operators in (\ref{threeeight}) would be
ordered by the value of $x$.

None of these probabilities occur in usual non-relativistic quantum
mechanics because that theory --- as defined in Section II ---  deals only with
alternatives at definite moments of the Newtonian time. To be consistent
with the usual theory, the operator $\hat t(\tau)$ defined above cannot mean
simply the time that the particle crosses $x$.
Although the classical trajectories of a free particle
cross a surface of constant $x$ at one time, non-relativistic quantum
mechanics predicts a probability for the particle to be at $x$ at {\it
any} time.  Put differently, in quantum mechanics the particle may cross
a surface of constant $x$ an arbitrary number of times.  There is no one
time of crossing for there to be a probability of. Some  interpretation,
beyond that of usual non-relativistic quantum mechanics, would have to
be given to the probabilities generated from the choice ${\cal T}=x$.

When there is any non-trivial dynamics, classical trajectories may
intersect a surface of constant $x$ more than once. For example if $V=
-gx$ then we can construct $\hat t(\tau)$ and $\hat p_x(\tau)$ by eliminating
$\lambda$ between
\begin{mathletters}
\label{fourtwo}
\begin{eqnarray}
\tau &=& \hat x + \frac{\hat p}{m} \lambda + \frac{1}{2}
g\lambda^2\ , \label{fourtwoa} \\
\hat t(\tau) &=& \hat t+\lambda\ , \label{fourtwob} \\
\hat p_x(\tau) &=& \hat p+g\lambda \ .
\label{fourtwoc}
\end{eqnarray}
\end{mathletters}
Thus, there would be {\it two} one-parameter families of operators
corresponding to the two branches in the solution to (\ref{fourtwoa}).
Again the resulting probabilities correspond to nothing predicted by
non-relativistic quantum mechanics.

The class of allowed time functions must be restricted for PNRQM to
coincide with usual quantum mechanics.  A possible restriction on time
functions of the form of ${\cal T}(t,x)$ would be to require that they
be ``good'' time functions
in the sense that all classical
trajectories intersect surfaces of constant ${\cal T}$ once and only
once. For many potentials this would restrict to ${\cal T}=t$.  Thus
restricted, PNRQM
would coincide with the usual theory.

Sch\"on and H\'aj\'\i\v cek \cite{Hajsum} have shown that good time
functions do not exist in generic minisuperspace models whose
constraints are analogous to those of general relativity. Their work
suggests that good time functions do not exist in the phase space of
general relativity. If that is the case, any restrictions on time
functions necessary for general relativity to coincide with usual
quantum mechanics in an appropriate limit would have be of a different
character.

\section{Time Functions and Spacetime}
\label{sec:V}

For many non-relativistic systems, good time functions of the form
${\cal T}(t, p_x)$ exist even when time functions depending only on
$t$ and $x$ are very
restricted. In this section we consider a simple
example of a time function ${\cal T} (t,p_x)$ and discuss its relations to the
usual
non-relativistic notions of spacetime.

Classically, $p_x$ will be a single-valued function of $t$ provided there
are no infinite forces, and for such  systems there will be good time
functions
of the form ${\cal T}(t,p_x)$. For example, $t+ kp_x$ is
a good time function for the free particle and simple harmonic
oscillator when $k$ is a constant of appropriate dimension and size. Consider
this time function in the
case of a free particle. Following the procedure of Section III, we may
identify the operators representing $x$ and $p_x$ when $t+k p_x=\tau$.
They are
\begin{mathletters}
\label{fiveone}
\begin{eqnarray}
\hat x(\tau) &=& \hat x + \frac{\hat p}{m} (\tau - k \hat p-\hat t)
\ , \label{fiveonea} \\
\hat p_x(\tau) &=& \hat p\ .
\label{fiveoneb}
\end{eqnarray}
\end{mathletters}

We would now like to describe a way in which measurements of quantity
like $\hat x(\tau)$ defined by (\ref{fiveonea}) differ in character from
those of $\hat x(t)$ where $t$ is the usual Newtonian time. First, note
that it is not enough simply to discuss or model the measurement of an operator
$\hat x(\tau)$ at some prescribed value of $\tau$ without also considering
how intervals of of $\tau$ are determined. That is because in
the Heisenberg
picture an operator represents {\it some} quantity at {\it
any} time. Consider by way of example the operator
\begin{equation}
\hat x + \frac{\hat p_x}{m} 5 \label{fivetwo}
\end{equation}
in the usual non-relativistic quantum mechanics of a free particle.
This represents $x$ at time $t=5$. But it could also represent a
particular combination of position and momentum at $t=0$. Similarly the
operator on the right hand side of (\ref{fiveonea}) represents $x$ when
$t+k p_x=\tau$ in the evolving constant scheme, but also represents a
combination of $x$ and $p_x$ when $t=\tau$, and, indeed, some other
combination for any other value of $t$.
In view of this, to model a measurement of $\hat x(\tau)$, one must not
only model a determination of an eigenvalue of the operator, but also
one must model a measurement of
intervals\footnote{\setlength{\baselineskip}{.1in} Strictly speaking,
the value of a time function itself cannot be measured since time
functions necessarily are not observables. However, {\it intervals} of
time functions can correspond to observables. For example, the integral
\begin{displaymath}
\int d\lambda (\dot t + k\dot p_x)
\end{displaymath}
can be reparametrization-invariant when carried out between
reparametrization-invariant end points such as might be defined by
positions of a clock indicator.} of the time parameter $\tau$.

The time function ${\cal T}=t$ is the unique one associated with the
Newtonian time defined by the geometry of non-relativistic spacetime.
Intervals of $t$ may be measured by a system (a clock) that is
completely independent of the free particle whose position is $x$ and
momenta $p_x$.
That is because usual
non-relativistic quantum mechanics assumes a fixed Newtonian spacetime
geometry and, in particular, the Newtonian idea of simultaneity.  The
readings of a clock may be simultaneous with the position of a particle
without there being any interaction between the two. Indeed, that is how
realistic measurements of quantities at near definite moments of time
are typically carried out.

By contrast,
it is difficult to see how the measurement of intervals of a
 time parameter like
$t+k p_x$ can be carried out by a clock that is independent of the particle
whose momentum is $p_x$. The $t$ in the time function may be the Newtonian
time, accessible to many different systems, composed of different
particles, but the $p_x$ refers to the {\it particular}
particle being measured.  A measurement of $x$ when $t+k p_x=\tau$
can certainly not be carried out by determining $x$, $p_x$ and $t$ at
one point along the particles trajectory as is shown by the numerous analyses
of
model experiments which attempt to simultaneously define $x$ and
$p_x$.
Intervals of time
functions such as $t+k p_x$,
that depend on the variables of that subsystem, cannot be
determined without intervention in the subsystem.

\section{Inconsistent Time Orderings}
\label{sec:VI}

\subsection{Particles}
\label{subsec:a}

In this section we show that PNRQM predicts inconsistent probabilities
for histories unless its allowable time functions are suitably
restricted.

We begin with a simple example in the PNRQM of a single particle
interpreted by the method of evolving constants, as described in Section
III. First, consider the time function ${\cal T}=t$ and the
alternatives that the particle is localized in a region of space
 $\Delta$ at a
value of the associated time parameter $\tau$, or not localized in that
region.  The alternative that the particle is in $\Delta$ when ${\cal
T} = \tau$ corresponds to the function on the reduced phase space
which is unity on those paths which cross $\Delta$ at that
time and zero otherwise. The alternative that the particle does not pass
through the region corresponds to the function which is zero on the paths
through $\Delta$ at $\tau$ and unity otherwise.

Evidently $\tau$ is the usual Newtonian time. As this is just
the usual quantum mechanical situation, there is no difficulty with
identifying the ``evolving constant'' corresponding to these alternatives. The
alternative that the particle is localized in $\Delta$ at $\tau$ is
represented by
\begin{equation}
\widehat P_\Delta (\tau)\ ,
 \label{sixone}
\end{equation}
where $\widehat P_\Delta(\tau)$ is the Heisenberg picture projection operator
onto the range $\Delta$ of $\hat x(\tau)$. The alternative that the
particle is not so localized is represented by
\begin{equation}
\skew3\widehat{\overline P}_\Delta (\tau) \equiv I -
\widehat P_\Delta(\tau)\ .
\label{sixtwo}
\end{equation}

Suppose measurements are carried out to determine whether the
particle is in a region $\Delta_1$ when ${\cal T} = \tau_1$ and in
$\Delta_2$ when ${\cal T} = \tau_2$ where $\tau_2 > \tau_1$. The
probability that the particle is so localized is, from (\ref{threeeight}),
\begin{equation}
p\left(\Delta_2, \Delta_1\right) = \Bigl\Vert \widehat P_{\Delta_2} (\tau_2)
\widehat P_{\Delta_1} (\tau_1) |\psi\rangle\Bigr\Vert^2\ ,
\label{sixthree}
\end{equation}
where the operators are ordered right to left according to increasing
values of the time parameter $\tau$.

\centerline{\epsfysize=5.00in \epsfbox{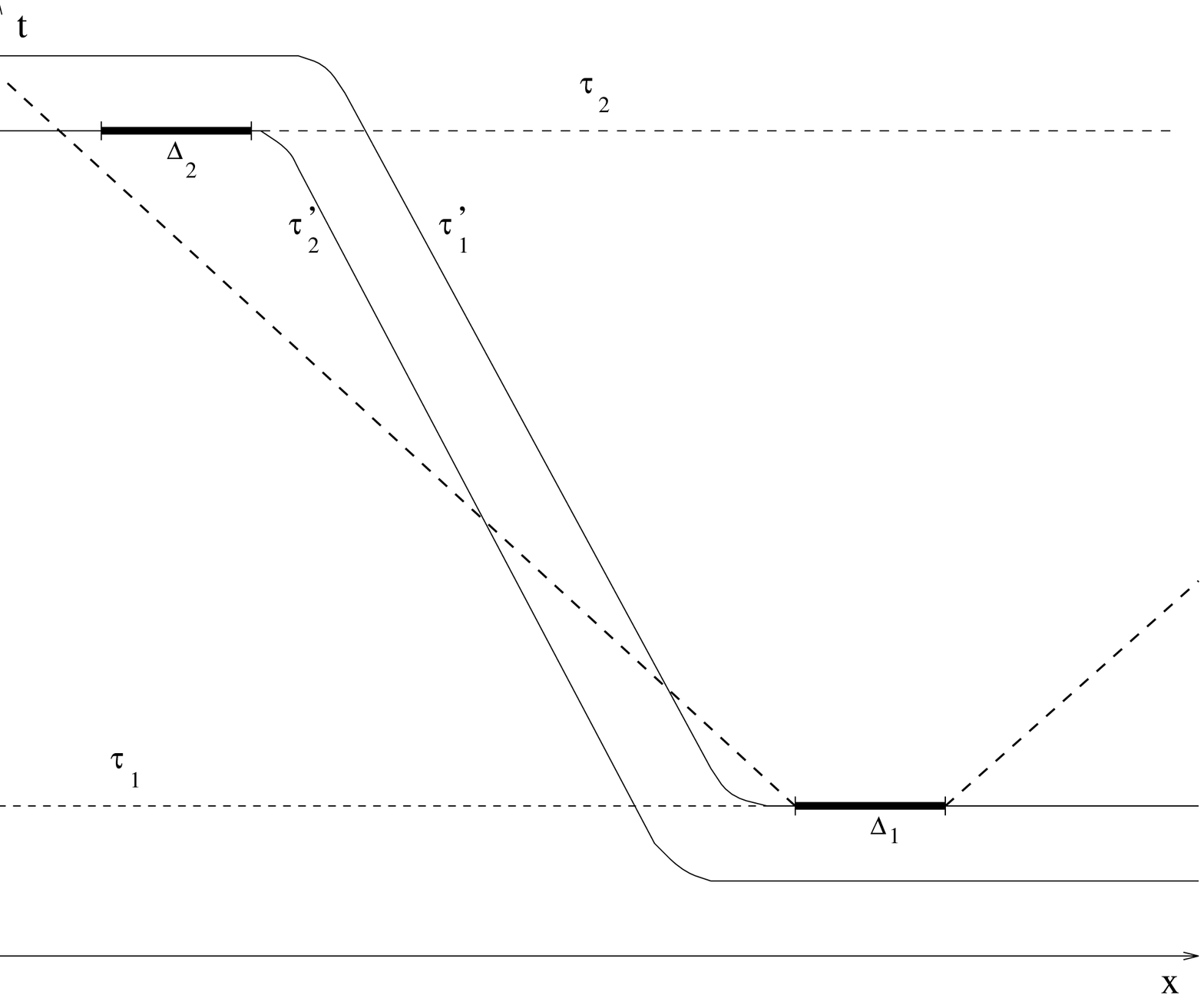}}
\vskip .26 in
{\setlength{\baselineskip}{.1in}
\it FIG 1: Inconsistent probabilities
for the same histories may arise
from two different choices of time function when the method of
``evolving constants'' is used with the usual quantum mechanical
framework for predicting these probabilities. The figure shows timelike
separated spacelike regions $\Delta_1$ and $\Delta_2$. (The diagonal
dotted lines
define the domain of causal dependence of $\Delta_1$.) The regions
$\Delta_1$ and $\Delta_2$ may be thought of as lying on the level
surfaces of a time function ${\cal T}(t, x)=t$ at constant  values
$\tau_1$  and
$\tau_2$ respectively. Alternatively they may be thought of as lying on
surfaces of
a time function ${\cal
T}^\prime(t,x)$ at constant values $\tau^\prime_1$ and $\tau^\prime_2$
shown by the solid lines above.
The operators representing alternatives that refer only
to $\Delta_1$ or $\Delta_2$ are independent of this choice, since the
level surfaces of the two time functions coincide in these regions.
However, the order ``in time'' of the alternatives is different in the
two cases.  Probabilities of histories depend on time order and will
therefore the two choices of time function will result in different
predictions for the probabilities of histories.
Consistency in this example could be restored by
restricting time functions to those whose level surfaces are spacelike ---
thus eliminating ${\cal T}^\prime$.}

\vskip .3in
Constant values of time functions that are functions of $t$ and $x$ alone
may be
thought of as defining surfaces in Newtonian spacetime. A region
$\Delta$ on a surface of constant ${\cal T}$ may equally well be
considered as a region on a surface of constant value
 of any other time function
 ${\cal T}^\prime(t,x)$ as long as the surfaces coincide inside $\Delta$.
(See Figure 1.) There are many ways ${\cal T}^\prime$ can differ from
${\cal T}$ outside $\Delta$ if no further restrictions are placed on
allowable time functions. The function on phase space which is unity on
the paths which pass through $\Delta$ at ${\cal T} = \tau$ and zero
otherwise is the same as the function which is unity on the paths which
pass through $\Delta$ when ${\cal T}^\prime = \tau^\prime$
and zero otherwise for
an appropriate $\tau^\prime$. The class of paths is the same in each case.
Assuming that the same operator ordering is used in both cases, the
corresponding ``evolving constant'' operators must coincide:
\begin{equation}
\widehat P_\Delta (\tau^\prime) = \widehat P_\Delta (\tau)\ .
\label{sixfour}
\end{equation}
This is only to be expected.
The operators
represent,  after all, the same alternative. Similarly there is an
equality between $\skew3\widehat{\overline P}_\Delta(\tau^\prime)$
and $\skew3\widehat{\overline P}_\Delta(\tau)$
following from (\ref{sixtwo}).

A problem of consistency arises when considering the probabilities of
histories of two such regions $\Delta_1$ and $\Delta_2$ at different
Newtonian times $\tau_1$ and $\tau_2$, $\tau_1<\tau_2$. As
illustrated in Figure 1, it is possible to choose another time function
${\cal T}^\prime(t,x)$ which is a surface of constant Newtonian time inside
$\Delta_1$ and $\Delta_2$, but such that the values of the time
parameters satisfy $\tau^\prime_1>\tau^\prime_2$.
Thus, the ordering of the same
alternatives is different according to the two choices of time function.
The probability $p(\Delta_1, \Delta_2)$ would be (\ref{sixthree}) using
Newtonian time ${\cal T}$ but
\begin{equation}
p(\Delta_1, \Delta_2) = \left\Vert \widehat P_{\Delta_1} (\tau^\prime_1)
\widehat P_{\Delta_2}
(\tau^\prime_2) | \psi\rangle\right\Vert^2
\label{sixfive}
\end{equation}
using the time function ${\cal T}^\prime$. Since the operators for the two
alternatives agree [cf.~(\ref{sixfour})], but do not commute, we have an
inconsistent assignment of probabilities to the same history.

A simple special case of the above ambiguity will serve to illustrate
explicitly
how the predictions of the ``evolving constant'' method can differ from those
of usual quantum mechanics if arbitrary time functions are allowed. Pick a
time function ${\cal T}^{\prime\prime}(t,x)$ corresponding to a family of
surfaces
in spacetime such that {\it both} region $\Delta_1$ at Newtonian time
${\tau}_1$ and $\Delta_2$ at Newtonian time $\tau_2$ lie on the {\it same}
surface ${\cal T}^{\prime\prime}
(t,x)=\tau^{\prime\prime}$. In the ``evolving constant''
construction the alternative that the particle passes through both
regions would then be represented by a {\it single} projection operator,
$\widehat P_{\Delta_1\Delta_2}(\tau^{\prime\prime})$. That projection operator
would be
constructed by taking the characteristic function on the region of
phase space corresponding to initial conditions $(x,p)_0$ for classical
trajectories that intersect both regions at the appropriate Newtonian
times and turning it into a projection operator with appropriate
ordering. Thus a {\it history} of alternatives which is represented by
a product of projections $\widehat P_{\Delta_2}(t_2)\widehat P_{\Delta_1}(t_1)$
in usual quantum mechanics would be represented by a {\it single}
projection in the ``evolving constant'' method using this choice of
time function.

This example points to another type of inconsistency which can arise
if the allowed time functions are not limited. Choose the time function
${\cal T}^{\prime\prime}(t,x)$ described above.
Physically, the alternative that the
particle lies in region $\Delta_1$ and $\Delta_2$ at time
$\tau^{\prime\prime}$ should
be the same as the history that the particle passes through $\Delta_1$
at time $\tau^{\prime\prime}$ and ``then'' passes through $\Delta_2$ at a time
$\tau^{\prime\prime} + \epsilon$ in the limit that $\epsilon$ tends to zero.
Yet, if we are correct that the operators representing the latter two
alternatives are independent of the choice of time function as in
(\ref{sixfour}), then the operators representing these two choices
would seem to be different. In the one case, the operator is a single
projection. In the other, it is a product of projections which do
not reduce to a single projection in the limit of vanishing $\epsilon$,
because they do not do so when represented as $\widehat P_{\Delta_2}(\tau_2)$
and
$\widehat P_{\Delta_1}(\tau_1)$. Indeed, unless the time functions are
suitably restricted, there will generally be an ambiguity in whether
a history of alternatives should be represented by a single operator that
is the transcription of the characteristic function of those points in the
reduced phase space corresponding to the {\it classical} trajectories
allowed by this sequence of alternatives, or by a sequence of operators
representing
a decomposition of the history into alternatives at different times,
whatever time function is chosen. Those two possibilities are not the
same because quantum mechanics permits non-classical trajectories.

The astute reader will have noticed that the time functions
${\cal T}^\prime$ and ${\cal T}^{\prime\prime}$
that lead to the above ambiguities, and to orderings different from that
provided by Newtonian time, are
not ``good'' time functions in the sense defined in Section IV. While
the classical trajectories that pass through $\Delta$ at ${\cal
T}^\prime=\tau^\prime$ intersect that surface once and only once,
the trajectories
defining the complementary alternative may intersect it many times.
Thus, restricting the theory to good time functions of the form ${\cal
T}(t,x)$ would eliminate this particular inconsistency. However, we
shall now see that
this kind of inconsistency persists, even for good time
functions, in  a theory that possesses a ``many-fingered''
parametrized time,

\subsection{Many-Fingered Times}
\label{subsec:b}

We consider a free massive, real relativistic field in one space and
one time dimension whose dynamics is summarized by the action
\begin{equation}
S[\phi(t,x)] = -\frac{1}{2}\int dt dx
\left[-\Bigl(\frac{\partial\phi}{\partial t}\Bigr)^2 +
\Bigl(\frac{\partial\phi}{\partial x}\Bigr)^2 + m^2 \phi^2\right]\ .
\label{sixsix}
\end{equation}
We foliate Minkowski spacetime by hypersurfaces
$t=t(\lambda, x)$ and regard the action as a
functional of both $\phi(\lambda, x)$ and $t(\lambda, x)$. The resulting
parametrized field
theory is invariant under independent reparametrizations of $\lambda$
for each $x$. There is therefore a constraint for each $x$ which has the
form
\begin{equation}
{\cal H}(x) \equiv \pi_t(x) + h\left[\pi_\phi(x^\prime), \phi(x^\prime),
t(x^\prime); x\right)
= 0\ .
\label{sixseven}
\end{equation}
Here, $\pi_\phi(x)$ and $\pi_t(x)$ are the momenta conjugate to
$\phi(x)$ and $t(x)$ respectively, and the bracket [ indicates that $h$
is a function{\it al} of the argument momenta and fields while the bracket
) indicates that it is a function of $x$.

The explicit form of $h$ is
\begin{equation}
h=\frac{1}{2}\left[(1-t^{\prime 2})^{-1} (\pi_\phi + t^\prime\phi^\prime)^2
+ (\phi^\prime)^2 + m^2
\phi^2\right]\ ,
\label{sixeight}
\end{equation}
where a prime indicates a partial derivative with respect to $x$.
The canonical form
of the action summarizing the resulting dynamics is
\begin{equation}
S\left[\pi_t, \pi_\phi, t, \phi\right] = \int d\lambda dx \Bigl\{\pi_t
\dot t + \pi_\phi \dot \phi - N {\cal H} \left[\pi_t, \pi_\phi, t,
\phi\right]\Bigr\}\ ,
\label{sixnine}
\end{equation}
where a dot denotes a partial derivative with respect to $\lambda$ and
$N(\lambda, x)$ is a multiplier enforcing the constraint.

The extended phase space of this parametrized field theory is
spanned by the co\"ordinates $\pi_t(x), \pi_\phi(x), t(x), \phi(x)$ ---
four co\"ordinates for each $x$. We denote these collectively by $z^A$
where, in the manner of DeWitt,
  the index $A$ indicates $x$ as well as whether the co\"ordinate is
$\pi_t, \pi_\phi, t$ or $\phi$. Since there is one constraint for each
$x$ the constraint surface has a correspondingly high co-dimension.

Classical motion in the constraint surface is generated by the
constraints (\ref{sixeight}) according to
\begin{equation}
\dot z^A(\lambda) = \Bigr\{z^A(\lambda), H[N]\Bigr\}\ ,
\label{sixten}
\end{equation}
where
\begin{equation}
H[N] = \int dx\, N(\lambda, x) {\cal H} \left[\pi_t, \pi_\phi, t, \phi;
x\right)
\label{sixeleven}
\end{equation}
for arbitrary $N(\lambda, x)$. Therefore, there is  not a unique {\it curve} or
trajectory describing classical evolution, but rather one function's worth of
{\it equivalent curves}
\begin{equation}
z^A = z^A(\lambda; N(\lambda^\prime, x^\prime)]\ .
\label{sixtwelve}
\end{equation}
This function's worth of equivalent
curves defines a point in the reduced phase
space.

We are now in a position to consider the generalization of our earlier
discussion of ``evolving constants'' to theories like this with an
infinite number of reparametrization
invariances. In the case of a single reparametrization invariance, we
defined  classical observables  at a point of the reduced phase space
(a classical trajectory) and
a value of a time parameter  by the
values of phase space functions at the point in the extended phase space
where the classical trajectory intersected a surface ${\cal
T}(z^A)=\tau$ specified by a time function ${\cal T}(z^A)$. However, the
function's worth of curves ({\ref{sixtwelve}) will intersect such a
surface in many points. A larger number of mutually intersecting time
functions is therefore needed to define a unique point where one
classical trajectory and all time functions coincide. One time function is
needed for each $x$. Thus we write
\begin{equation}
{\cal T}\bigl[z^A; x\bigr) = \tau(x)\ .
\label{sixthirteen}
\end{equation}
There is a time function for each $x$ and $\tau(x)$ is its value. Thus
there is not just one time parameter but one for each $x$ --- a freely
specifiable ``many-fingered time''. Classical observables are
functionals of $\tau(x)$.  The operator ordering problems are even more
formidable in this environment, but, assuming that they can be solved, we
write $\widehat F[\tau(x)]$ for an evolving constant operator satisfying
\begin{equation}
\Bigl[\widehat F[\tau(x^\prime)], \widehat {\cal H}(x)\Bigr] = 0\ .
\label{sixfourteen}
\end{equation}
Similarly projections onto ranges of values of $\widehat F$ are labeled by
$\tau(x)$.

We now turn to the ordering inconsistencies that can arise in such a
framework when defining the probabilities of histories. First it is
clear that there is no universal notion of the ordering of functions
$\tau_1(x)$ and $\tau_2(x)$ in the same way that there is for single
parameters $\tau_1$ and $\tau_2$. However, one could consider
restricting the notion of history to sequences of $\tau(x)$'s for which
\begin{equation}
\tau_n(x) > \tau_{n-1}(x) > \cdots > \tau_1(x)\ ,
\label{sixfifteen}
\end{equation}
for each $x$. While the physical content of such a restriction is obscure in
general to this author, a meaning can be given for that
limited class of time functions which correspond to hypersurfaces in
spacetime.  We now restrict attention to this class.

The time function most closely related to the time of spacetime geometry
is
\begin{equation}
{\cal T}\bigl[z^A; x\bigr) = t(x)\ .
\label{sixsixteen}
\end{equation}
The values
\begin{equation}
t(x) = \tau(x)
\label{sixseventeen}
\end{equation}
of this time function
 define surfaces in spacetime. Histories are naturally restricted to
sequences of $\tau(x)$'s which are members of a foliation of spacetime
and which therefore satisfy (\ref{sixfifteen}).

The time function $t(x)$ is a ``good'' time function. Fields
$\phi(\lambda, x)$ are single valued on spacetime, so for a given
$\tau(x)$ there is only one value for $\phi, \pi_\phi, t$ and $\pi_t$ on
the surface $t(x) = \tau(x)$. Yet, as we shall now show, ordering
ambiguities of the kind discussed for particles in Section \ref{subsec:a}
remain if
$\tau(x)$ corresponding to all possible foliations of spacetime by
hypersurfaces are permitted.

Consider the two timelike separated regions $\Delta_1$ and $\Delta_2$ as
shown in Figure 1. We may consider alternative values of quantities
constructed only from the values of fields in these regions. For instance, we
could consider the average value of the field in one region. Such
alternatives may be said to be at the time parameters $\tau_1(x)$ and
$\tau_2(x)$ which are surfaces of constant $t$. Alternatively they may be
said to be at the time parameters $\tau^\prime_1(x)$ and
$\tau^\prime_2(x)$  which
are constant time surfaces inside $\Delta_1$ and $\Delta_2$ but vary
outside. The operators corresponding to the alternatives are independent
of which of the two sets of time parameters is used since they are the
same alternatives in either case. However, the ordering of operators in
the construction of the probability of
 a history is different depending on which of the two
assignments of many-fingered time is used. Further, the two orderings
are inconsistent since the operators will not commute because the regions
$\Delta_1$ and $\Delta_2$ are timelike separated. If they are
spacelike separated then the ordering is ambiguous, but it is also
irrelevant since fields at spacelike separated points commute.

A restriction to ``good'' time functions will not eliminate this
inconsistency as it did for the case of the particle in Section
 \ref{subsec:a}.
The time
function $t(x)$ is a good time function for fields as discussed above.
Instead, to express usual field theory in this language, one would need
to restrict, not the time function itself, but rather its values
$\tau(x)$ so that they correspond to foliations of spacetime by {\it
spacelike} surfaces. This restriction eliminates the surfaces
$\tau^\prime_1(x)$ and $\tau^\prime_2(x)$ since they necessarily contain some
non-spacelike parts if $\Delta_1$ and $\Delta_2$ are timelike separated.

\subsection{General Relativity}
\label{subsec:c}

With suitable restrictions on the allowed time functions, the method of
``evolving constants'' can yield a consistent quantum theory of
particles and fields in a fixed background spacetime.
Restrict to ``good'' time functions. Restrict to time functions that
correspond to surfaces in the background spacetime. Restrict to
values of many-fingered time that define foliations of that fixed spacetime
by spacelike surfaces.
The result is the usual quantum mechanics of particles and
fields in fixed background spacetimes. But what of such restrictions in
theories of quantum general relativity where spacetime geometry
is not fixed? We
have already mentioned that there may be no ``good'' time functions for
general relativity. Certainly there cannot be a restriction to time
functions that define surfaces in spacetime in a theory where spacetime
geometry is not fixed. The closest analogy would be time functions
that define surfaces in the superspace of three-geometries. In
superspace many different curves, corresponding to different ways of
foliating a classical spacetime, describe the same classical evolution.
The issue of consistency
of the order of operators defining histories will therefore certainly
arise.  It seems unlikely that it can be resolved by restricting to
``spacelike'' surfaces in superspace. The DeWitt metric gives a natural
notion of spacelike directions in superspace, but there is unlikely to
be any notion of causality in superspace similar to that in field theory.
Even classical trajectories propagate outside the light cone of
superspace. Further, some popular choices of time functions, such as
York's trace of the extrinsic curvature, do not even define surfaces in
the superspace of three geometries. For such reasons,
the nature of the restrictions on time functions and
on time parameter values that would give a consistent ``evolving
constant'' quantum dynamics of spacetime geometry coinciding with the
usual fixed background theories in an appropriate limit
remains an unsolved problem.

\section{Conclusions}
\label{sec:VII}

Parametrized non-relativistic quantum mechanics (PNRQM) interpreted by
the method of evolving constants does not coincide with the
usual non-relativistic quantum theory unless the class of
allowable time functions is severely restricted. To be sure, PNRQM
reproduces the predictions of the usual theory for the choice of time
function ${\cal T}(t, x, p_x) = t$. However, to the extent that it allows
other time functions it makes predictions that go beyond those of
familiar non-relativistic theory. At best these new predictions may be
generalizations of the usual framework presenting new challenges for
interpretation and understanding. Certainly the central role played by
a fixed Newtonian spacetime geometry in the usual theory is altered. At
worst, for some non-relativistic systems, the predictions for histories
 may be
inconsistent  unless the time functions are suitably restricted or understood
by
a new interpretative rule.

There is no evidence of experiment or theoretical principle compelling a
generalization of non-relativistic quantum mechanics such as PNRQM
interpreted by the method of evolving constants represents. However, in
the opinion of many workers, including the author, generalization of the
familiar quantum mechanical framework, closely tied as it is to a notion of
fixed background geometry, is a natural route to resolving the ``problem
of time'' in quantum gravity. If the principles of any such
generalization are to apply to the non-relativistic domain to yield a
{\it necessary} generalization of the usual theory, then we are indeed in a
fortunate position. For then the generalization can be analyzed for
logical and experimental consistency in a well understood situation that
is much
more accessible to theoretical computation and experimental test than
any involving quantum gravity. At present it is probably not clear
whether the method of evolving constants applied to a putative quantum
theory of gravity implies that a PNRQM of particles or fields should
apply in the non-relativistic domain. If so, however, progress
in understanding can be made by resolving the issues of meaning and
consistency raised above and demonstrating consistency with known
physics. That is an obligation which is equally shared by other proposed
generalizations of usual quantum mechanics.

\acknowledgments

The author thanks C.~Rovelli for extensive correspondence which has been
helpful in clarifying the method of evolving constants. Thanks are also
due to J.~Halliwell, K.~Kucha\v r, D.~Marolf, and W.~Unruh for useful
conversations. Special thanks are due to S.~Carlip, P.~H\'aj\'\i\v cek,
K.~Kucha\v r, D.~Marolf, and C.~Rovelli for critical readings of the
manuscript.
This work was supported in part by the NSF under grant PHY90-08502.

\end{document}